\documentclass{article}
\usepackage{graphicx} 
\usepackage{hyperref}
\usepackage{amsfonts}

\usepackage{amsmath}
\usepackage{mathtools}

\usepackage[top=0.5in, bottom=0.7in, left=0.8in, right=0.8in]{geometry}

\title{Predicting Plasma Temperature From Line Intensities Using ML Models}
\author{Ashwini Malviya \\ \textit{Department of Physics} 
\textit{IIT Roorkee}\\
Roorkee, India \\
a\_malviya@ph.iitr.ac.in}

\date{}

\begin{document}
\maketitle
\begin{abstract}
\textbf{
In this work, ML models were used to predict the plasma temperature using the dataset obtained by implementing the CR-model for Na-like Krypton.  The models included in the study are: Linear Regression, Lasso Regression, Support Vector Regression, Decision Trees, Random Forest, XGBoost, Multi-layer Perceptron and Convolutional Neural Network. For evaluating the models we used Mean Absolute Error, Mean Squared Error and $R^2$ Score as metrics, In our study Random Forest performed  best as compared to other model considered, the study conclude that complex relation between the line-intensities and Plasma temperature can be capture by ML models and they can be used to predict the temperature with high accuracy.}
\end{abstract}
\section{Introduction}

The study of plasma of Na-like (Isoelectronic with Na) Argon, Krypton and Xenon is important in fusion reactors as these noble gas ions are used present as impurity seeding\cite{Beiersdorfer_2015}. Plasma is most abundant state of matter in outer space we makes its study  importance in astrophysics. The discovery of hot Argon ion in interstellar space highlights the significance of these ions in plasma studies \cite{refId0}. Keeping that in mind various studies have been performed to accurately calculate the atomic parameter \cite{Hill2022SummaryRO},\cite{PhysRevA.102.062823,PhysRevA.55.3229,article, kr}, one such method is the  Collision-Radiative (CR)\cite{JOHNSON2019100579,HARTGERS2001199} modeling of plasma. The calculation involves the Multi
configuration Dirac–Hartree–Fock (MCDHF)\cite{grant2007relativistic} and Relativistic
 Distorted Wave (RDW)\cite{GUPTA2020106992} theories that provides the wave function, fine structure energies, electron-impact-excitation cross-section and other quantities required for CR modeling. With CR model  we can calculate the emission line intensities as a function of plasma temperature. It requires knowledge of quantum mechanics and atomic physics to perform such calculation.  In this work we used Machine learning (ML) models to predict the plasma temperature based on the line-intensity data. Specifically we used Classical regressions, tree-based and deep learning models.  

 \section{Theory}
 In multiconfiguration Dirac-Hartee-Fock theory, the  linear combination of Configuration state functions (CSFs) provides the atomic state function (ASF). CSFs are given by an antisymmetric product of single electron Dirac orbitals.
\begin{equation}
\Psi(\gamma PJM) = \sum_{j=1}^{n_C} c_j \Phi(\gamma_j PJM)
\end{equation}
where $\psi$ is ASF, \textit{P, J, M} are parity, total angular momentum, and its projection. $\phi$ is CSF, and $c_j$ is the mixing coefficient, representing the corresponding CSFs' contribution. 

For calculating the single electron orbitals Dirac-Hartee-Fock equations are used the  Dirac-Coulomb Hamiltonian can be represented as: 
\begin{equation}
\hat{H}_{DC} = \sum_{i=1}^{N} \left[ c (\boldsymbol{\alpha} \cdot \mathbf{p}_i) + (\beta - I)c^2 + V_i \right] + \sum_{i<j} \frac{1}{r_{ij}}
\end{equation}
where c, is speed of light, $\alpha, \beta$ are Dirac matrices, $p_i$ momentum operator,$V_i$ represent potential energy, and $\frac{1}{r_{ij}}$ is coulumb repulsion. MCDHF uses variational principle to minimize the total energy which is given by 
\begin{equation}
E[\Psi] = \frac{\langle \Psi | \hat{H}_{DC} | \Psi \rangle}{\langle \Psi | \Psi \rangle}
\end{equation}
After calculating the wavefunctions from MCDHF theroy, RDW theory can used to calculate the trasntion matrix which is given by 

\begin{equation}
T_{l \to u}^{\text{RDW}}(J_u M_u, \vec{k}_u \mu_u ;  J_l M_l, \vec{k}_l \mu_l, \theta)=
     \langle \phi_{u}^{rel}(1,2,\dots ,N) F_{u,\mu_u}^{DW-}(\vec{K_u},N+1)|V-U_f| 
 \times A \phi_{l}^{rel}(1,2,\dots,N)F_{l,u_l}^{DW+}(\vec{k_l},N+1) \rangle \hspace{20pt}
 \end{equation} 
 
 \vspace{10pt}
 
here $\phi$ represent bound state wave functions ( $l$ and $u$ denotes lower and upper state) , $J$ and $M$ are total angular momentum and corresponding magnetic component.
 $F_{l(u),\mu_{l(u)}}^{DW+} $ is a relativistic distorted wave function for incoming (outgoing) electron having wave vector $\vec{k_{l(u)}}$ having magnetic components $\mu_{l(u)}$, the angle between the incident and scattered electron is given by $\theta$
 and $-(+)$ denotes incoming(outgoing) electron boundary condition.

 With the help of transition matrix one can calculate the parameters required. 
These parameter can be used to build CR-model.
In CR-model the kinetic equations are solved for the population of levels. In this work we included following process in kinetics. 

Electron impact excitation and de-excitation:

\begin{equation}
    A_i + e + Q \longleftrightarrow A_f + e 
\end{equation}

Electron impact ionization and three-body recombination: 
\begin{equation}
    A_p + e + (Q_{p+}) \longleftrightarrow X_{+}+ e + e 
\end{equation}

Spontaneous radiative decay:
\begin{equation}
    A_p \longleftrightarrow A_q + hv_{pq}
\end{equation}

where $A, X$ represent atom while $e$ is for electron, further $Q$ and $hv$ are energies.  

\subsection{ML models}

\subsubsection*{Linear Regression (LR)}
In Linear Regression  model the relationship between a scalar dependent variable $y$ and one or more explanatory variables $x$ is given by:
\begin{equation}
y = \beta_0 + \beta_1 x_1 + \beta_2 x_2 + \cdots + \beta_p x_p + \epsilon
\end{equation}
where: $y$ is the response variable,$x_i$ are the predictors,$\beta_i$ are the coefficients,$\epsilon$ is the error term.
The parameters $\boldsymbol{\beta}$ are estimated by minimizing the residual sum of squares (RSS):
\begin{equation}
    \min_{\boldsymbol{\beta}} \sum_{i=1}^n \left( y_i - \mathbf{x}_i^\top \boldsymbol{\beta} \right)^2
\end{equation}

\subsubsection*{Ridge Regression}

Ridge Regression is a regularized version of Linear Regression that adds an $L_2$ penalty to the loss function. It helps to prevent overfitting by shrinking the regression coefficients.

The objective function minimized in Ridge Regression is:

\begin{equation}
\min_{\boldsymbol{\beta}} \left\{ \sum_{i=1}^{n} (y_i - \mathbf{x}_i^\top \boldsymbol{\beta})^2 + \lambda \sum_{j=1}^{p} \beta_j^2 \right\}
\end{equation}

This can also be written in matrix form as:

\begin{equation}
\min_{\boldsymbol{\beta}} \left\{ \| \mathbf{y} - \mathbf{X} \boldsymbol{\beta} \|_2^2 + \lambda \| \boldsymbol{\beta} \|_2^2 \right\}
\end{equation}

where:
$\mathbf{X} \in \mathbb{R}^{n \times p}$ is the design matrix,
    $\mathbf{y} \in \mathbb{R}^n$ is the target vector,
    $\boldsymbol{\beta} \in \mathbb{R}^p$ is the coefficient vector,
    $\lambda \geq 0$ is the regularization parameter.
The closed-form solution is given by:

\begin{equation}
\boldsymbol{\beta} = (\mathbf{X}^\top \mathbf{X} + \lambda \mathbf{I})^{-1} \mathbf{X}^\top \mathbf{y}
\end{equation}

where $\mathbf{I}$ is the identity matrix of size $p \times p$.

\subsubsection*{Lasso Regression}
Lasso (Least Absolute Shrinkage and Selection Operator) is a linear regression technique that performs both variable selection and regularization. It minimizes the following cost function:
\begin{equation}
\min_{\boldsymbol{\beta}} \left\{ \sum_{i=1}^n \left( y_i - \mathbf{x}_i^\top \boldsymbol{\beta} \right)^2 + \lambda \sum_{j=1}^p |\beta_j| \right\}
\end{equation}
where: $\lambda \geq 0$ is a regularization parameter, the $L_1$ penalty $\sum |\beta_j|$ promotes sparsity in the coefficients.

\subsubsection*{Support Vector Regression (SVR)}
SVR aims to find a function $f(x) = \mathbf{w}^\top \phi(x) + b$ that approximates the targets within an $\varepsilon$-tube, minimizing both the model complexity and the deviation outside the tube.

The optimization problem is:
\begin{equation}
\min_{\mathbf{w}, b, \xi, \xi^*} \frac{1}{2} \|\mathbf{w}\|^2 + C \sum_{i=1}^n \left( \xi_i + \xi_i^* \right)   
\end{equation}
subject to:
\[
\begin{aligned}
y_i - \mathbf{w}^\top \phi(x_i) - b &\leq \varepsilon + \xi_i \\
\mathbf{w}^\top \phi(x_i) + b - y_i &\leq \varepsilon + \xi_i^* \\
\xi_i, \xi_i^* &\geq 0
\end{aligned}
\]
where: $\phi(x)$ maps $x$ into a higher-dimensional space,
    $\varepsilon$ is the tube size,
    $C$ is the regularization parameter,
    $\xi_i$, $\xi_i^*$ are slack variables for the margin violations.

\subsubsection*{Decision Trees}

Decision Trees partition the feature space recursively to minimize the prediction error. At each split, the algorithm selects the feature and threshold that minimizes a loss function, such as Mean Squared Error (MSE) for regression:

\begin{equation}
\text{MSE} = \frac{1}{n} \sum_{i=1}^{n} (y_i - \hat{y}_i)^2
\end{equation}

The optimal split at a node is found by:

\begin{equation}
\min_{j, s} \left[ \frac{n_L}{n} \text{MSE}_L + \frac{n_R}{n} \text{MSE}_R \right]
\end{equation}

where:$j$ is the feature index,
    $s$ is the split point,
    $n_L$, $n_R$ are the number of samples in left/right nodes,
    $\text{MSE}_L$, $\text{MSE}_R$ are the MSEs in left/right nodes.

\subsubsection*{Random Forest}

Random Forest is an ensemble method that constructs multiple decision trees and outputs the average prediction for regression tasks:

\begin{equation}
\hat{y} = \frac{1}{T} \sum_{t=1}^{T} f_t(x)
\end{equation}

where: $T$ is the number of trees,$f_t(x)$ is the prediction from the $t$-th tree.

Each tree is trained on a bootstrap sample and uses a random subset of features at each split to reduce correlation between trees.

\subsubsection*{XGBoost (Extreme Gradient Boosting)}

XGBoost is a gradient boosting algorithm that builds trees sequentially to correct the residual errors of the previous trees. The model is:

\begin{equation}
\hat{y}_i = \sum_{k=1}^{K} f_k(x_i), \quad f_k \in \mathcal{F}
\end{equation}

The objective function is:

\begin{equation}
\mathcal{L}(\phi) = \sum_{i=1}^{n} l(y_i, \hat{y}_i) + \sum_{k=1}^{K} \Omega(f_k)
\end{equation}

where the regularization term is:

\begin{equation}
\Omega(f) = \gamma T + \frac{1}{2} \lambda \sum_{j=1}^{T} w_j^2
\end{equation}
 $l$ is a loss function (e.g., squared error),
    $\Omega(f)$ penalizes model complexity,
    $T$ is the number of leaves in a tree,
    $w_j$ is the score on leaf $j$.

\subsubsection*{Multi-Layer Perceptron (MLP)}

An MLP is a feedforward neural network composed of an input layer, one or more hidden layers, and an output layer. Each layer applies an affine transformation followed by a nonlinear activation function.

The output of a hidden layer is:

\begin{equation}
\mathbf{h}^{(l)} = \sigma \left( \mathbf{W}^{(l)} \mathbf{h}^{(l-1)} + \mathbf{b}^{(l)} \right)
\end{equation}

where:
$\mathbf{h}^{(l)}$ is the output of layer $l$,
    $\mathbf{W}^{(l)}$ and $\mathbf{b}^{(l)}$ are the weight matrix and bias vector of layer $l$,
    $\sigma$ is a nonlinear activation function (e.g., ReLU, sigmoid).

For the output layer in a regression setting:

\begin{equation}
\hat{y} = \mathbf{W}^{(L)} \mathbf{h}^{(L-1)} + \mathbf{b}^{(L)}
\end{equation}

The model is trained by minimizing a loss function $L(y, \hat{y})$ over the dataset.

---

\subsubsection*{Convolutional Neural Network (CNN)}

CNNs are specialized neural networks designed to process data with grid-like topology (e.g., images). A convolutional layer applies a set of filters to extract spatial features.

For a 2D input image $X$ and filter $K$, the convolution operation is defined as:

\begin{equation}
S(i,j) = (X * K)(i,j) = \sum_m \sum_n X(i+m, j+n) \cdot K(m,n)
\end{equation}

After convolution, an activation function $\sigma$ is applied:

\begin{equation}
A(i,j) = \sigma(S(i,j))
\end{equation}

A CNN typically consists of multiple layers including:
Convolutional layers (feature extraction),
    Pooling layers (dimensionality reduction),
    Fully connected layers (classification/regression).
    
The network is trained by backpropagation using a loss function, often optimized with gradient descent.

\section{Dataset}

In this study we  considered the line-intensities of Na-like Krypton. The data was obtained by implementing CR model for Na-like Krypton, the process considered (also deceibed in previous section) in the CR model are: Electron-impact excitation/de-excitation, Electron impact ionization/Three-body recombination and Spontaneous radiative decay. Finally we calculate the intensity by solving the kinetics equations. We obtained four  transitions lines  correspond to $2p^63p( \prescript{2}{}{P_{1/2}^{o}})-2p^63s(\prescript{2}{}{S_{1/2}})$,  $2p^63d( \prescript{2}{}{D_{3/2}})-2p^63p(\prescript{2}{}{P_{3/2}^{o}})$, $2p^63d( \prescript{2}{}{D_{5/2}})-2p^63p(\prescript{2}{}{P_{3/2}^{o}})$, and $2p^63d( \prescript{2}{}{D_{3/2}})-2p^63p(\prescript{2}{}{P_{1/2}^{o}})$ transitions which are at 22.00,17.89,15.99, and 14.08 nm respectively in the emission spectra. The intensities of these lines is function of plasma temperature therefore we used the normalized intensities for these lines  as input to ML modes to  predict the plasma temperature. So each example consist of four features (normalized intensities of four transitions) and one output (Temperature). In total our dataset consist of 19500 examples in which temperature ranges from 500 ev to 20000 ev. The dataset along with other resouces can be found at \href{https://github.com/ash-007-m/PREDICTING-PLASMA-TEMPERATURE-FROM-LINE-INTENSITIES-USING-ML-MODELS}{GitHub Repository}.

\section{Result and Discussion}
In this section we provide the results for considered ML models: 
 \vspace{1em}
 \newline
 1. Classical: Linear Regression(LR),Ridge Regression, Lasso Regression and Support Vector Regression(SVR). \newline
 2. Tree-Based: Decision Trees, Random Forest and XGBoost.\newline
 3. Deep Learning models: Multi-Layer Perceptron (MLP) and Convolutional Neural Network (CNN).

 \vspace{0.5em}
 For model evaluation we used: Mean Absolute Error (MAE), Root Mean Squared Error (RMSE), R² Score (coefficient of determination) as performance Metrics.

\subsection{1. Classical models}
\begin{figure*}[h!]
  \centering
 \includegraphics[width=16.5cm, height=10.8cm]{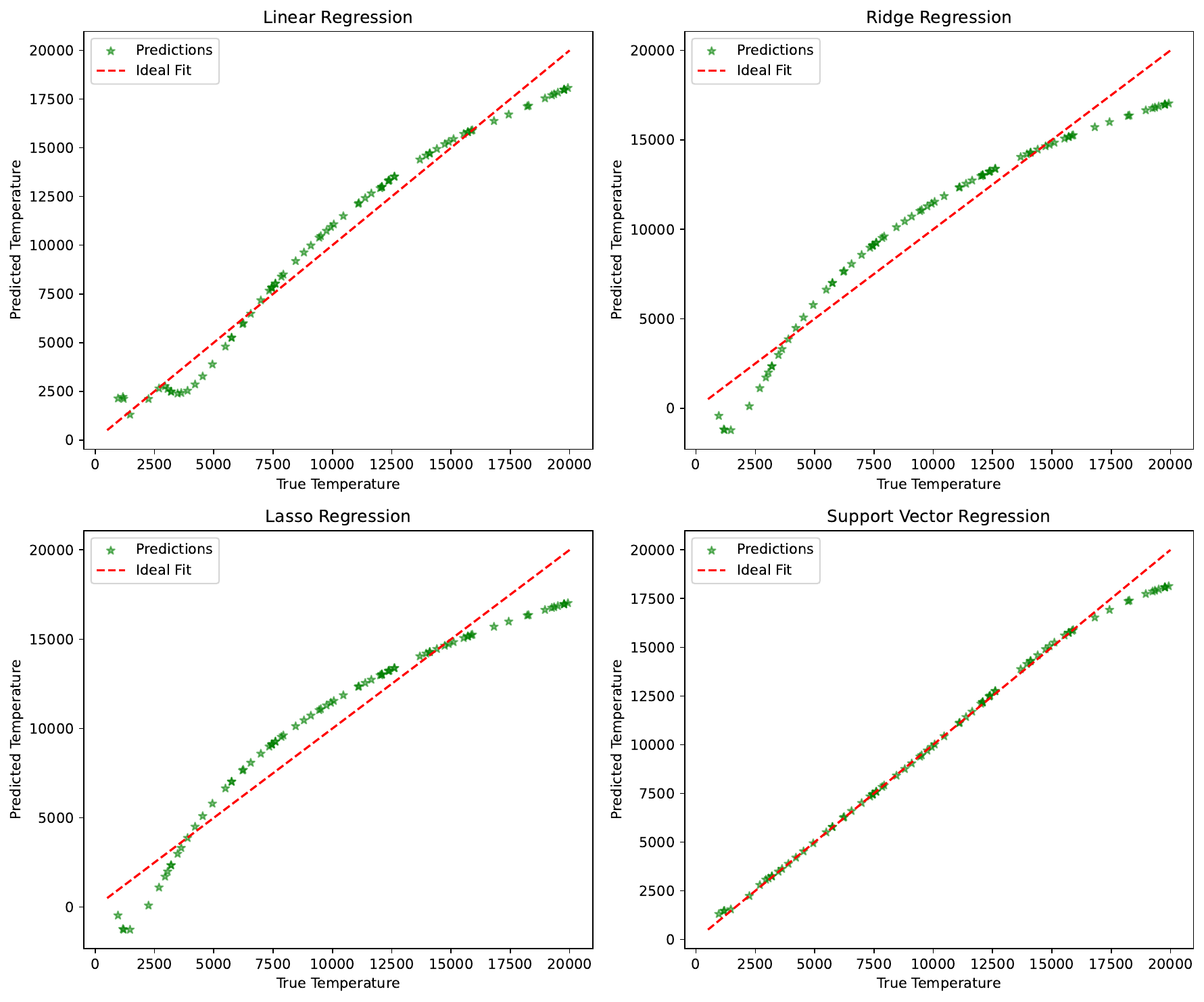} 

  \caption{Prediction for classical models}
 \label{classical}
 \end{figure*}

Table \ref{classical table} provides the results for Linear, Ridge, Lasso, and Support Vector Regression. 
The train-test ratio  was 80:20 and we used 5-fold Cross validation to avoid overfitting and set the hyperparameter. The $'rbf'$ kernel was used for SVR. We see that all these models have large MAE, which means that they are not able to capture the dependence of temperature on line-ratios. Fig\ref{classical} shows the predicted temperature vs true temperature for these models, the red-dashed line shows the ideal fit. We can see that SVR performed better for the Temperature(T) in middle range while for lower and higher temperature it shows large deviations, whereas other regressions have small deviations for every section(low, mid and high T). So among classical models we find that SVR perform best although it has high MAE and RMSE overall. 

 \begin{table*}[h!]
    \centering
    \caption{}
    \hspace{2em}
    \label{classical table}
    
    \begin{tabular}{c c c c}
    \hline
   Model &  MAE  &      RMSE & R² Score \\
        \hline
        \hline
    Linear Regression &          746.241014  & 860.169377 & 0.976207\\ \hline
Ridge Regression &          1281.272664  &1491.187458  &0.928495\\ \hline
Lasso Regression   &        1292.112345 & 1503.423491  &0.927316\\ \hline
Support Vector Regression &  249.325521&   526.095265 & 0.991100\\
   \hline
    \end{tabular}
\end{table*}

\subsection{2. Tree-based model}
 
 \begin{table*}[h!]
    \centering
    \caption{}
    \hspace{2em}
    \label{tree}
    
    \begin{tabular}{c c c c}
    \hline
   Model &  MAE  &      RMSE & R² Score \\
        \hline
        \hline
Decision Tree &  1.040000 &  1.058785 & 1.000000\\ \hline
Random Forest &  0.451159 &  0.559367 & 1.000000\\ \hline
XGBoost(Gradient Boosting)      & 18.158805 & 21.383741 & 0.999985\\ \hline

    \end{tabular}
\end{table*}

  \begin{figure*}[h!]
  \centering
 \includegraphics[width=16.5cm, height=6.8cm]{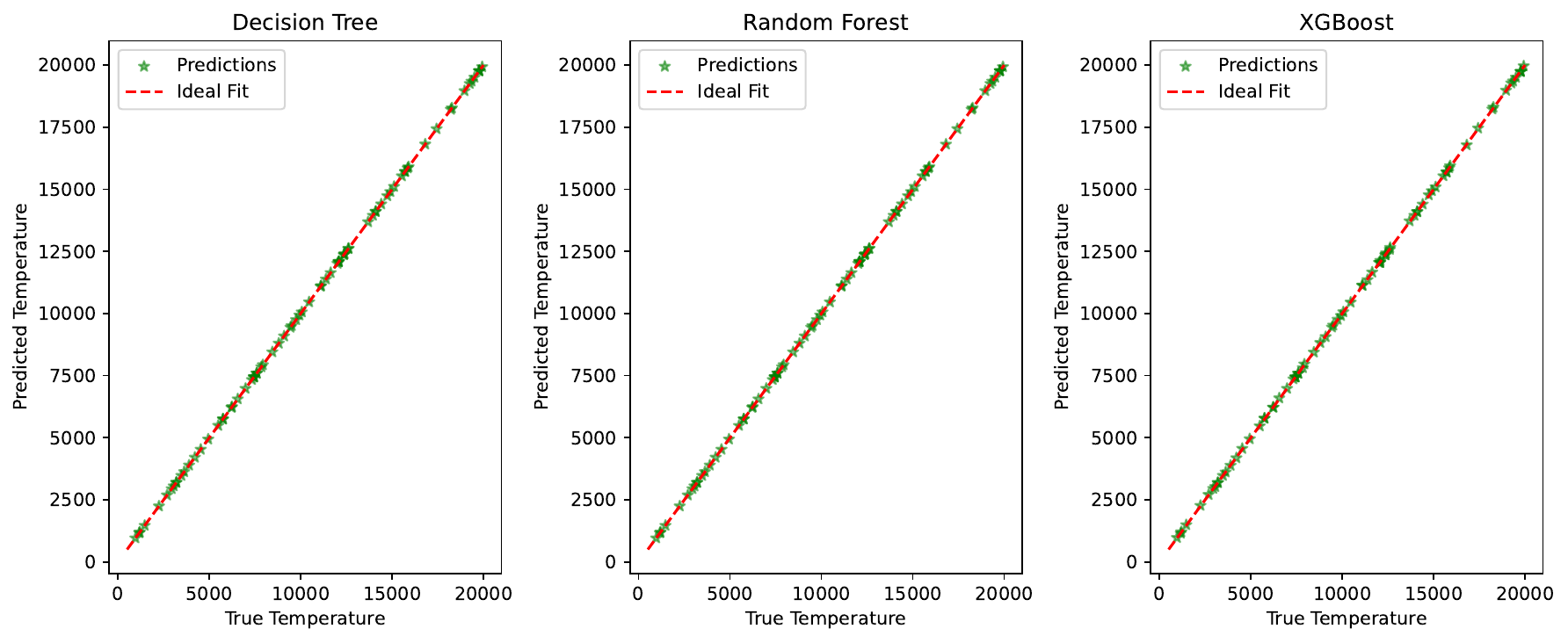} 
 \caption{Prediction for Tree-based model}
 \label{tree fig}
 \end{figure*}

 Table \ref{tree} provides the results for Decision tree, Random Forest and XGBoost. Again we used 80:20 ratio for training-testing and 5-Fold Cross validation. Tree-based model performed better than regression models(and overall), the MAE is least for Random Forest. These models were able to capture the non-linearity present in the data so could perform better than regression models.
 
 Fig \ref{tree fig} also shows that the predicted value is very close to the true value. Further among the tree-based model we can see that Random Forest is the best for our prediction as we can see from the MAE, RMSE and $R^2$ value, Decision tress are Basic model and  prone to overfitting, whereas Random forest being the ensemble of trees reduce variance, usually XGBoost gives better generalization but on our dataset it did not perform better than Random Forest. Note that we used hyperparameter tuning for both Random Forest and XGBoost in which we considered: Number of trees in the forest, Maximum depth of each tree, Minimum number of samples to split a node, Minimum number of samples in a leaf for tuning Random forest and Number of boosting rounds, Step size shrinkage, Depth of trees and Fraction of samples used per boosting round for setting XGBoost parameters.  

 \subsection{3. Deep learning models}
\begin{figure*}[h!]
  \centering
 \includegraphics[width=11.5cm, height=5.0cm]{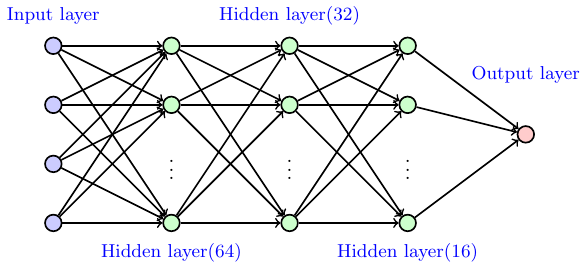} 
 \caption{Multilayer-Perceptron architecture}
 \label{MLP}
 \end{figure*}

\begin{figure*}[h!]
  \centering
 \includegraphics[width=15.5cm, height=2.0cm]{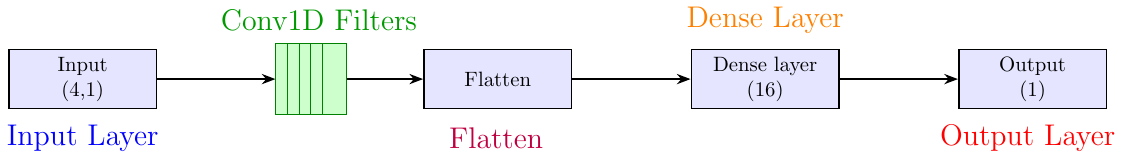} 
 \caption{Convolutional Neural Network architecture}
 \label{CNN}
 \end{figure*}

 Multilayer-perceptrons (MLP) and Convolutional Neural Network (CNN) are included in case of deep learning models. CNN is knows for capturing the spatial relations between input. we used fully connected
 network for MLP, the architecture is shown in Fig\ref{MLP}, we have 4 line intensities as input then we used 3 hidden layer having 64, 32 and 16 nodes. For each node we used 'relu' activation function. 
 For CNN we have $4 \times 1$ as input layer and we do a 1D convolution, which followed by flattening , and then dense layer, again we use 'relu' activation function  for output. This simple architecture is shown in Fig\ref{CNN}.

\begin{figure*}[h!]
  \centering
 \includegraphics[width=16.5cm, height=5.8cm]{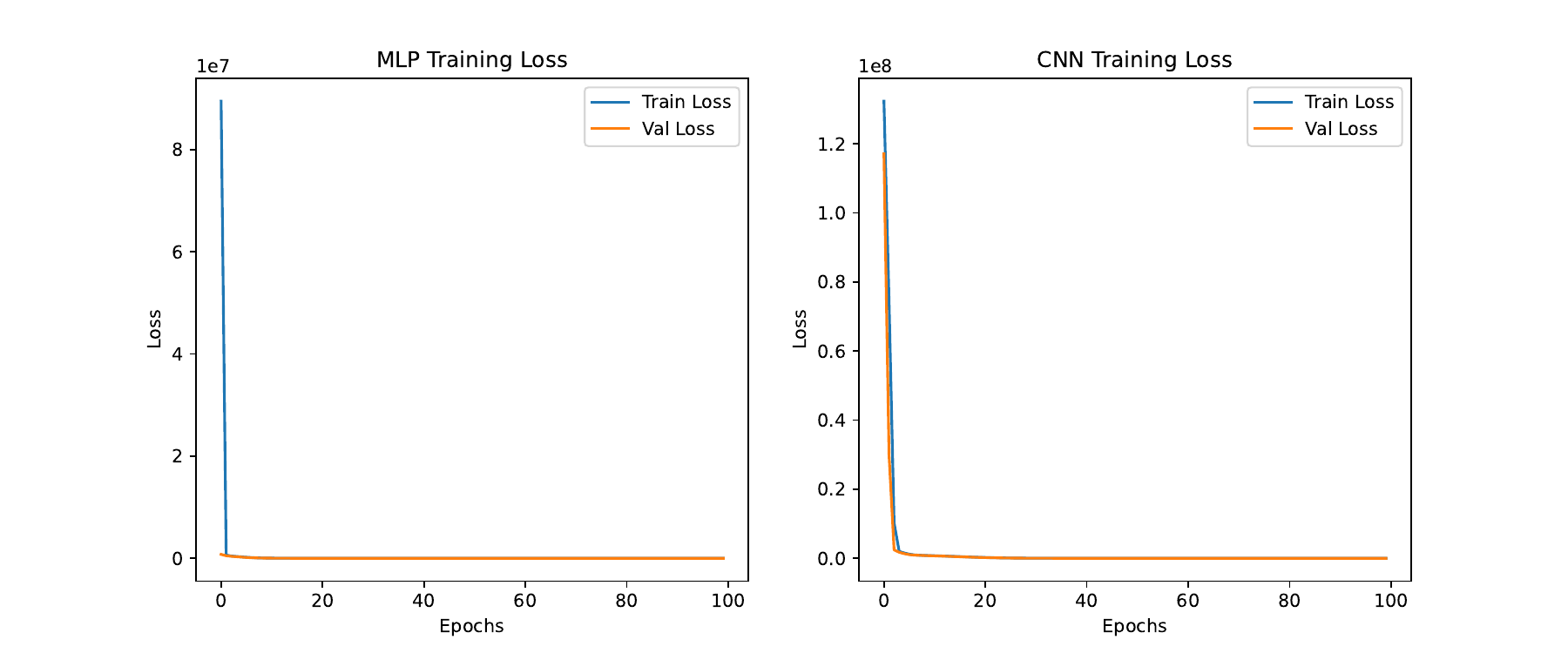} 
 \caption{Prediction for Tree-based model}
 \label{deep fig}
 \end{figure*}

 For both the models used 'adam' optimizer and 'MSE' as error function, and Keras Tuner was used to tune the hyperparameter.
 The results for these two models are summarized in Table \ref{deep table}.These model have low MAE and RMSE than regression models, but have more error as compared to tree-based models. Fig\ref{deep fig} provides the validation loss and training loss as function of epochs. we see that the result converges within few epochs only, although we considered simple architecture for CNN, but its able to capture the relation between input and output variables as performed better than MLP. 
 \begin{table*}[h!]
    \centering
    \caption{}
    \hspace{2em}
    \label{deep table}
    
    \begin{tabular}{c c c c}
    \hline
   Model &  MAE  &      RMSE & R² Score \\
        \hline
        \hline
Multi-Layer Perceptron (MLP) &  8.429787  & 12.987405 & 0.999995 \\ \hline
Convolutional Neural Network (CNN)&  4.689976  & 8.333597 & 0.999998 \\ \hline

    \end{tabular}
\end{table*}

\section{Conclusion}
The study conclude that the complex dependence of line-intensities on Temperature can captured by the ML models and they can be used in the studies related to plasma modeling. In our analysis we found that Random Forest performed best over considered other models. Although we have discussed simple cases, but the results motivates us to study more complex system and to investigate other start-of-the art ML models and methods. 
\bibliographystyle{IEEEtran} 
\bibliography{Mybib.bib}
\end{document}